# Electrical Nanoprobing of Semiconducting Carbon Nanotubes using an Atomic Force Microscope

Y. Yaish, J.-Y. Park, S. Rosenblatt, V. Sazonova, M. Brink, P. L. McEuen

*Laboratory of Atomic and Solid-State Physics, Cornell University, Ithaca, NY 14853*

We use an Atomic Force Microscope (AFM) tip to locally probe the electronic properties of semiconducting carbon nanotube transistors. A gold-coated AFM tip serves as a voltage or current probe in three-probe measurement setup. Using the tip as a movable current probe, we investigate the scaling of the device properties with channel length. Using the tip as a voltage probe, we study the properties of the contacts. We find that Au makes an excellent contact in the p-region, with no Schottky barrier. In the n-region large contact resistances were found which dominate the transport properties.

PACS #'s: 71.20.Tx, 73.63.Fg, 73.40.Cg

Semiconducting carbon nanotubes (NTs) show great promise as field-effect transistors with device properties that can exceed those of Si MOSFETs [1-3]. Furthermore, NTs are very robust, operate under a variety of conditions, and are compatible with many other materials and fabrication techniques. As a result, they are being actively investigated for a number of applications, from high performance electronics to molecular sensing [4-7].

One point of controversy has been the role played by the metallic contacts to semiconducting NTs. These contacts have been studied extensively theoretically [8-12] and experimentally [13-16]. Theoretically, it is expected that the contacts will either be Schottky barriers (SB) or Ohmic depending on the work function difference between the metal and the NT and the type of carrier (n or p) (Figure 1(b)-(d)). Experiments by some groups [3, 13, 17, 18] have been analyzed assuming that this contact resistance is low, as in a traditional MOSFET. Other experiments [14, 15, 19-21], however, provide evidence that transport is dominated by Schottky barriers at the contacts. Resolving this issue is critical for probing the ultimate limit of NT transistors.

In this paper, we directly measure the contact resistance in semiconducting NTs using a new technique where a metallized AFM tip serves as a scanned electrical nanoprobe (SEN). We show that, for Au contacts and p-type device operation, Schottky barriers are not formed and that contact resistances approach theoretical limits. In the n-region, on the other hand, we observe large resistances near the metal-tube interface that are attributable to barriers formed at the interface between the p-type contacts to the n-type device.

The devices used in this study, shown schematically in Fig 1(a), are made by CVD growth [22] of NTs at lithographically defined catalyst sites on Si/SiO$_2$ substrate with oxide thickness of t = 200nm or 500nm. The electrodes were made by evaporation of either 5nm Cr/50nm Au or 30 nm Au on top of the NTs followed by liftoff [3]. The doped substrate is used as a gate to change the carrier density in the tube.

We first consider device operation in the p-region. Semiconducting tubes with Cr/Au contacts typically showed low conductances (G ~ 0.001-0.1 $e^2/h$) in the on state, but annealing at 600°C for 10 minutes in an Argon environment dramatically improved the on-state conductance (G ~ 0.1-1.5 $e^2/h$). Devices made without the Cr adhesion layer showed high on-state conductances without annealing.

To locally probe their properties, the samples are placed in a commercial AFM system (Dimension 3100, Digital Instruments) operating in air at room temperature [23]. The three-probe measurement technique employed is shown in Fig. 1(a). The contacts labeled as 1 and 2 provide two of the probes, while the third probe is the metallized AFM tip. The tip is positioned above the tube at a specific location and then lowered until it makes physical and electrical contact with the NT. Previous work has shown [24] that a Au-coated tip can make very good electrical contact to a NT, with the tip-tube resistance $R_{tip-tube}$ < 100 kΩ.

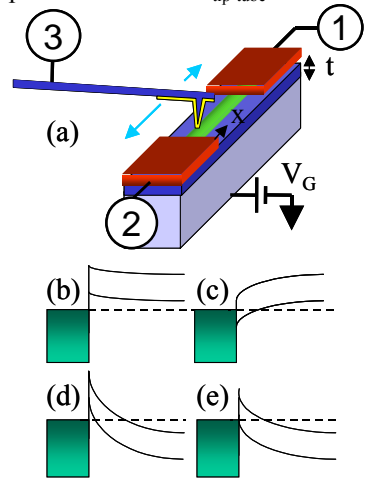

Fig 1 (color online). (a) A schematic of the three-probe measurement setup. The two electrodes and the AFM tip can each serve as a voltage source, current probe, or voltage probe. (b), (c) Band diagram in the p-region for Ohmic and Schottky contacts, respectively. (d), (e) Band diagram in the n-region for Ohmic and Schottky contacts.

We begin by using the SEN as a source current contact, as shown schematically in the inset to Fig. 2. Current flows from the SEN to the drain contact. The main panel shows the two-terminal conductance G vs. gate voltage $V_G$ for different values of the channel length $x$ for a NT with diameter d = 2 nm. As the channel length is decreased, the on-state conductance grows. At the shortest length (20nm) the maximum on-state conductance is ~1.3 $e^2$/h. This is within a factor of 3 of the theoretical limit of $4e^2$/h for a NT with a single channel occupied, indicating that both the contact and channel resistances are very low [25].

Since the NT/tip contact resistance can change for each measurement, we cannot quantitatively separate the channel and contact contributions from the data in Fig. 2. To determine them, a three terminal measurement using the SEN as a local voltage probe is employed. A voltage $V_{SD}$ is applied between the source and drain, producing a current through the device. The local voltage at specific locations along the tube is measured using the SEN by a voltmeter attached to the AFM tip as it is repeatedly touched down at a specific point on the tube [26].

Fig. 3 shows the voltage $V$ measured at various places along a 1.2μm long, 3nm diameter NT for $V_{SD}$ = 100 mV and for two $V_G$'s corresponding to total device resistances $R_T$ of 50 kΩ and 330 kΩ. For $R_T$ = 50 kΩ, the potential drop along the tube is linear, with significant voltage drops at the contacts. From the measured voltage drops, we infer that the tube intrinsic resistance $R_{tube}$ ~ 20 kΩ and the source and drain contact resistances are $R_S$ ~ $R_D$ ~ 15 kΩ. Near turn off ($R_T$ = 330 kΩ) most of the voltage drop occurs near the center of the tube, with only a small amount at the contacts: $R_{tube}$ ~ 300 kΩ, and $R_S$ ~ $R_D$ ~ 15 kΩ.

The inset to the figure shows $R_T$ and $R_S$ (measured 70 nm from the Au electrode) as a function of $V_G$. $R_T$ changes by an order of magnitude, but $R_S$ is relatively constant. Similar results have been obtained on other semiconducting tubes. For example a d = 1.4 nm tube had relatively constant contact resistances of $R_S$ ~ 30 kΩ and $R_D$ ~ 50 kΩ, while $R_T$ changed by an order of magnitude.

We can also directly measure the contact resistance between the NT and the AFM tip by using the SEN as a current probe, as in Fig. 2, and using one of the contacts as a voltage probe. We routinely find resistances in the range of 10-30 kΩ. The tip radius is small (~ 20 nm), indicating that even a small Au contact can make good contact to a p-type NT.

These results unambiguously demonstrate that Au can make very good contact to a p-type semiconducting nanotube, with typical resistances $R_{S, D}$ ~ 10- 50 kΩ. This resistance does not depend on $V_G$, unlike in a Schottky barrier [15]. Furthermore, these contact resistances are comparable to, but still somewhat higher than, the theoretical limit of $h/8e^2$ = 3.2 kΩ per contact.

The measurements above also allow us to find the intrinsic resistivity of a p-type nanotube in the on-state, where the voltage drop is approximately linear. From the data of Fig. 3, $R_{tube}/L$ = 20 kΩ/μm. Using $R_{tube} = h/4e^2 \cdot L/\ell_0$ the mean free path is then $\ell_0 \approx 300 nm$. The shortest channel devices in Fig. 2 are thus likely ballistic, showing why the conductance does not change significantly for channel lengths below 200 nm. The longer devices are still in the diffusive regime with the conductance decreasing with channel length.

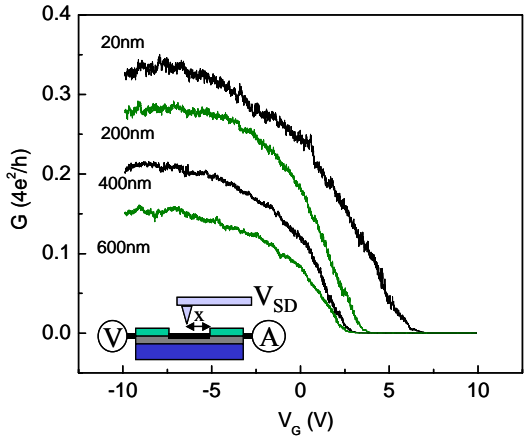

Fig 2 (color online). Conductance vs. gate voltage, for different distances x between tip and drain (x = 20, 200, 400 and 600nm). $V_{tip}$= 100mV, d = 2nm, and t = 500nm. Inset: Measurement configuration. The source-drain bias $V_{SD}$ is applied to the tip, while the drain (right) electrode is connected to current amplifier, and the left electrode to a voltmeter.

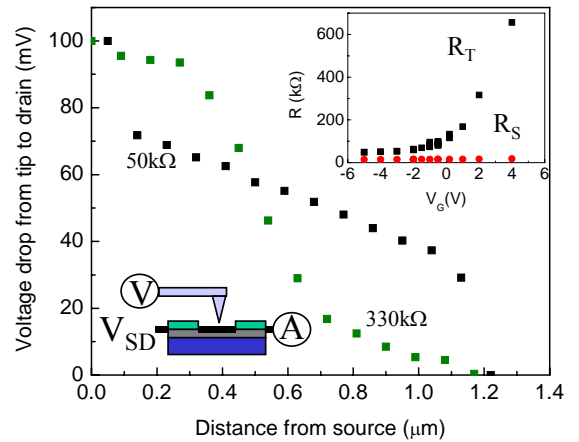

Fig 3 (color online). Profile of the voltage drop along tube with L = 1.2μm, d = 3nm, and $V_{SD}$ = 100mV. The two curves are for two different gate voltages, which correspond to total resistance $R_T$ of 50kΩ, and 330kΩ. Upper right inset: $R_T$ and source contact resistance $R_S$ (measured 70nm from lead) vs. gate voltage. Lower left inset: measurement configuration.

To verify the results above, we must check the invasiveness of the SEN when used as a voltage probe. In general, the presence of the tip will change the properties of the device, both by locally changing the electron density and also by inducing additional scattering. This will lead to errors in the measured voltage, particularly on length scales less than the mean free path. In the diffusive limit, however, the resistances of different parts of the tube add linearly, and the size of these errors can be estimated. In this limit, if the tip causes a resistance change $\delta R$ in the device, it will lead to an error in the measurement of the voltage of $\delta V/V \sim \delta R/R$. In the lower inset of Fig. 4 we plot a time trace of the total current and the local voltage measured as the SEN is brought into contact with the tube, for different $V_G$'s. The change in the current is less than 5% for two cases shown in the p-region ($V_G = -1$ V, $R_T = 60$ k$\Omega$ and $V_G = 1.2$ V, $R_T = 0.9$ M$\Omega$). This is typical for all our measurements, indicating that the voltage probe is not significantly invasive in the p-region.

When the tube is n-type, however, changes 10-30% are common; a decrease of 30% is seen in the data of Fig. 4 ($V_G = 5$ V, $R_T = 1.6$ M$\Omega$). In the n region, therefore, the probe is significantly invasive and voltage measurements are only qualitatively valid. The invasiveness of the AFM tip is likely because the p-type Au contact creates a barrier in the n-type tube.

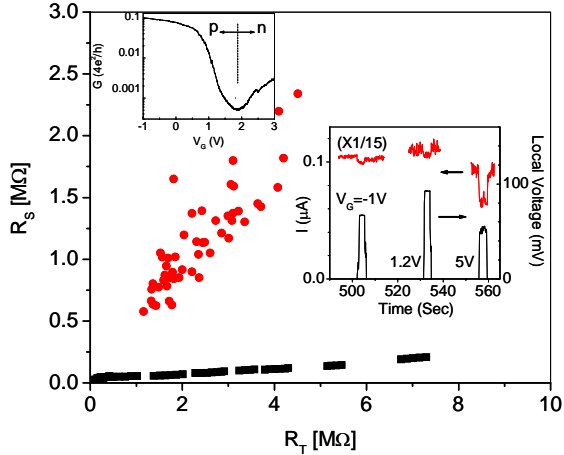

Fig. 4 (color online). Source contact resistance $R_S$ (measured 100nm from lead) vs. total resistance $R_T$ for $V_{SD} = 100$mV, L = 1.4$\mu$m, and d = 4nm. Squares correspond to p-doped and circles to n-doped regions. The slight increase of $R_S$ with $R_T$ in the p region is likely due to the resistance changes of 100nm segment of the NT between the source and the tip. Upper inset: G vs. $V_G$, plotted on a log scale. The p and n regions are shown. Lower inset: I-time (left axis), and local voltage vs. time (right axis), for three different $V_G$'s as the tip is brought into contact with the tube. $V_{SD} = 100$mV and for $V_G = -1$V, the current is divided by 15.

With the above limitation in mind, we examine contact resistance measurements in the n-type region. The upper inset to Fig. 4 shows the total conductance $G_T$ vs. $V_G$, plotted on a log scale, for a NT of length 1.4$\mu$m and d=4nm. The conductance in the n-region is much smaller than in the p-region [16, 27]. The main panel shows the source contact resistance $R_S$ (100nm from the source electrode), plotted versus the total device resistance $R_T$. In the p region $R_S$ is relatively constant while in the n-region it increases dramatically and is roughly half of the total resistance. The drain contact resistance gives similar results. Most of the resistance of the device is associated with the regions near the contacts in the n-region. Measurements of other devices yield similar behavior.

These results are consistent with previous experiments, where large resistances observed in the n-region were ascribed to either Schottky barriers or p-n junctions near the contacts [19, 20, 27-30]. Au makes p-type contacts to the tube, pinning the Fermi level in the valence band at the tube/Au interface (see Fig. 1(d)). Furthermore, the metallic contacts screen the effect of the gate on the tube within a distance $\sim t$ of the contacts [9]. This leads to barriers for transport for n-type operation. The screening effect is clearly seen in the data of Fig 2. As the channel length $x$ decreases, threshold for p-type conduction shifts to more positive $V_G$ when $x$ becomes smaller than the oxide thickness $t$.

We end by studying the relation between these experiments and previous works, where evidence for Schottky barriers at the contacts in the p-region have been found [14, 16, 17, 19-21, 29]. We observe similar effects in devices with Cr/Au contacts before annealing. An example is shown in Fig. 5, for a NT of length 6$\mu$m and $d$ = 2nm. In the main panel we plot the $I$-$V_{SD}$ curves for the device in the p-region before and after annealing. Before annealing the low bias resistance is high ($R_T \sim 30$M$\Omega$) and the $I$-$V$ is strongly rectifying. We use the SEN as a voltage probe to measure the local voltage drop at the drain (right) contact (Fig. 5.). The drain barrier is highly resistive and shows rectifying behavior, characteristic of a Schottky barrier. The source barrier is much more conductive, but also shows evidence of Schottky barrier behavior at larger biases (not shown). After annealing, the device resistance is much lower ($R_T \sim 200$k$\Omega$), the $I$-$V$ characteristics are almost linear and the contact resistances have decreased dramatically to 20-30k$\Omega$.

We attribute the Schottky barriers seen in Fig. 5 to the presence of the 5nm Cr layer. The Au work function is higher by 0.6V than Cr, so the Au Fermi energy coincides with states in the tube valence band (Fig. 1(b)) while the Cr Fermi level lies in the bandgap of the NT (Fig. 1(c)). During annealing, we postulate that the Au penetrates the thin Cr layer to create a direct Au contact to the tube.

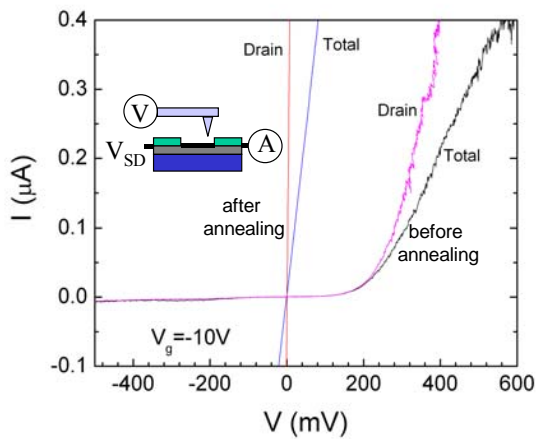

Fig 5 (color online). Current-Voltage curves for a device with Cr-Au contacts before and after annealing. The voltage V refers to either the total voltage $V_{SD}$ across the device or the voltage measured by the AFM tip 50nm from the drain lead. Inset: Measurement configuration.

The choice of metals is therefore critical to the nature of the metal/NT contact. For example, the IBM group found SB behavior using electrodes made of Ti [19], which has a work function similar to Cr. They also found the resistance increased dramatically upon cooling due to suppression of thermionic emission over the SB [21]. To test for small SBs in our Au-contacted devices, we cooled them to 10K. The conductance increased on cooling, further confirming that no Schottky barrier is present.

In conclusion, we have used an Au-coated AFM tip as a scanning electrical nanoprobe to investigate NT transistors. This method enables us to measure separately the resistances between the tube and the contacts, the resistance between the tube and the tip, and also the intrinsic resistance of the tube itself. We find that Au makes an excellent p-type contact to the tube, without measurable Schottky barriers. Measurements in the n-region show large contact resistances.

We acknowledge useful conversations with Hongjie Dai. This work was supported by the NSF Center for Nanoscale Systems, the Packard Foundation, and the MARCO/DARPA FRC-MSD, which is funded at MIT under contract 2001-MT-887 and MDA972-01-1-0035. Sample fabrication was performed at the Cornell node of the National Nanofabrication users Network, funded by NSF.